\documentclass[twocolumn,10pt,journal]{IEEEtran}

\ifCLASSINFOpdf
\else
\fi

\usepackage{longtable}
\usepackage{amsmath}
\usepackage{cancel}
\usepackage{amssymb}
\usepackage{graphicx}
\usepackage{longtable}
\usepackage{multirow}
\usepackage{array}
\usepackage[labelsep=period]{caption}
\usepackage{setspace}
\usepackage{float}
\usepackage{subfig}
\usepackage{cite}
\usepackage{cleveref}
\usepackage{color}
\usepackage{balance}

\newenvironment{remark}[1][Remark]{\begin{trivlist}
		\item[\hskip \labelsep {\bfseries #1}]}{\end{trivlist}}

\newcommand{\qed}{\nobreak \ifvmode \relax \else
	\ifdim\lastskip<1.5em \hskip-\lastskip
	\hskip1.5em plus0em minus0.5em \fi \nobreak
	\vrule height0.75em width0.5em depth0.25em\fi}

\begin{document}
%
\title{Outage Analysis of Cognitive Electric Vehicular Networks over Mixed RF/VLC Channels}
\author{Galymzhan~Nauryzbayev,~\IEEEmembership{Member,~IEEE,}
	Mohamed~Abdallah,~\IEEEmembership{Senior~Member,~IEEE,} 
	and Naofal Al-Dhahir,~\IEEEmembership{Fellow,~IEEE}

\thanks{G. Nauryzbayev is with the Department of Electrical and Computer Engineering, School of Engineering and Digital Sciences, Nazarbayev University, 010000, Nur-Sultan city, Kazakhstan (e-mail: galymzhan.nauryzbayev@nu.edu.kz).}
\thanks{M. Abdallah is with the Division of Information and Computing Technology, College of Science and Engineering, Hamad Bin Khalifa University, Qatar Foundation, Doha, Qatar (e-mail: moabdallah@hbku.edu.qa).}
\thanks{N. Al-Dhahir is with the Department of Electrical and Computer Engineering, University of Texas at Dallas, TX 75080, Dallas, USA (e-mail: aldhahir@utdallas.edu).}%

	}

\maketitle

\thispagestyle{empty}
\begin{abstract}
	\boldmath
	Modern transportation infrastructures are considered as one of the main sources of the greenhouse gases emitted into the atmosphere. This situation requires the decision-making players to enact the mass use of electric vehicles (EVs) which, in turn, highly demand novel secure communication technologies robust to various cyber-attacks. Therefore, in this paper, we propose a novel jamming-robust communication technique for different outdoor cognitive EV-enabled network cases over mixed radio-frequency (RF)/visible light communication (VLC) channels. One EV acts as a relaying node to allow an aggregator to reach the jammed EV and, at the same time, operates in both RF and VLC spectrum bands while satisfying interference constraints imposed by the primary network entities. We derive exact closed-form analytical expressions for the outage probability and also provide their asymptotic analysis while considering various channel state information quality scenarios. Moreover, we quantify the outage reduction achievable by deploying such mixed VLC/RF channels. Finally, analytical and simulation results validate the accuracy of our analysis. 
\end{abstract}

\IEEEpeerreviewmaketitle

\section{Introduction}
Current transport infrastructures are regarded as a significant source of the carbon dioxide (${\rm CO}_2$) gases emitted into the atmosphere which consequently increase the global temperature. For instance, it was noticed that the transport-based ${\rm CO}_2$ emission comprises $27 \%$ of the entire greenhouse impact in the United States \cite{USreport}. Therefore, a smooth transition from a carbon-greedy society to a carbon-free one can sufficiently improve this situation. Known for being highly efficient with low ${\rm CO}_2$ emission and better performance, electrical vehicles (EVs) have recently received significant public attention as an effective way to reduce the overall level of greenhouse gases. On the other hand, mass exploitation of EVs imposes several requirements on the current power grid (PG) systems such as being able to generate, aggregate and distribute large amounts of low-carbon energy in a real-time manner. 

Such an integration of the EVs and PG is enabled by means of a reliable and secure bi-directional communication networks. Therefore, several communication technologies ({\it i.e.}, Zigbee, WiFi, WiMAX, power-line communication (PLC) and cognitive radio (CR) networks) have been applied to support an interaction between the PG and individual EVs \cite{USpaper,MCOMM}. For instance, Zigbee is a low-cost solution that can provide a satisfactory rate of 250 kbit/s and mostly suitable for intra-vehicle communications. At the same time, this technology has some drawbacks such as low delay tolerance and vulnerability to the engine noise and radio-frequency (RF) interference \cite{IOTjournal}. Moreover, the authors in \cite{naofal1,naofal2} derived closed-form bit-error rate expressions and investigated the diversity of the proposed hybrid PLC/RF technique for the PG and vehicle-to-grid communications. 

Integrating the PG with different communication technologies not only improves the efficiency and flexibility of the PG but also makes the latter vulnerable to cyber security risks typical to these communication systems \cite{MCOMM,v2g}. In general, these cyber security threats are usually divided into four categories: authenticity, integrity, confidentiality, and availability \cite{cyber_threats}. The latter is the most important one to maintain communication services. For example, the loss of availability does not allow the PG to dynamically respond to the real-time power requests. On the other hand, the EVs which are disconnected from the PG will experience poor charging services. With the public information in the commercial wireless systems, illegal nodes can jam or distort the RF signal at the destination by transmitting an interfering signal over the air. As a result, the jammed or distorted received signal becomes unrecognizable. Such jamming attacks performed against the critical PG node can disable the operation of a large area since a central service controller does not obtain accurate load information. Consequently, local load unbalancing can affect a broader area \cite{MCOMM}.

The exponentially growing number of wireless devices and massive EV exploitation naturally result in RF spectrum scarcity which can be handled by using the CR paradigm \cite{Haykin,tcom2,galymTVT2020,sultanTCCN}. CR is an emerging wireless technology allowing an unlicensed user, known as a secondary user (SU), to use a licensed spectrum band such that a primary user (PU) does not experience harmful interference caused by the SUs \cite{sultanTCCN}. Such communication can be organized using three known approaches, {\it i.e.},  interweave, underlay and overlay, which have different operational modes \cite{Goldsmith}. For instance, the underlay CR grants permission to the SUs to operate in the licensed spectrum given that this communication does not result in excessive interference power at the PUs, which is referred to as an interference temperature constraint (ITC) \cite{sultanCOMML1}. Moreover, it is worth mentioning that the CR features perfectly match the mass EV exploitation. In general, the EVs are highly dynamic and therefore can connect to the PG at any site irrespective of the type of a certain local communication network. Regarding the jamming attacks, the conventional anti-jamming techniques, {\it e.g.}, frequency-hopping spread spectrum, direct-sequence spread spectrum, {\it etc.}, can not be applied to the CR network since they do not take into account the unavoidable impact of the presence of primary network entities \cite{MCOMM,cr_v2g}.

Recently, visible light communications (VLC) has emerged as an attractive candidate to resolve the RF spectrum scarcity and avoid the CR restrictions simultaneously \cite{galymPhotJ,galymAccess,PTL,RicianVLC}. The main advantages of VLC technology are its electromagnetic immunity to the RF interference and the availability of a wide unregulated visible light spectrum range (from 430 THz to 790 THz) used for concurrent illumination and communication purposes. This technology can be applied to the indoor and outdoor environments. For example, the authors in \cite{PTL} investigated the outage performance of the indoor non-orthogonal multiple access based VLC network scenario. In \cite{RicianVLC}, it was experimentally demonstrated that the outdoor VLC channels follow the Rician statistical distribution.

In this paper, we considered the downlink outdoor electric vehicular network when one EV is used as a relay to communicate with another EV whose direct link to the aggregator is jammed. We also specify several EV network scenarios to fully investigate the outage performance. Hence, our contributions are as follows. First, new closed-form analytical expressions were derived for the outage probability (OP) in outdoor cognitive relaying networks over mixed RF-RF/VLC channels. Second, we quantify the outage reduction enabled by deploying such a hybrid VLC/RF communication link. Third, we asymptotically analyze the OP and the corresponding outage reduction using the high signal-to-noise ratio (SNR) approximation and perfect CSI scenario. Forth, the provided analysis is validated by a good agreement with Monte Carlo simulations.

The remainder of the paper is organized as follows. In Section \ref{sec:system_model}, we describe the system model while, in Section \ref{sec:analysis}, we derive the closed-form expressions for the OP and outage reduction of the considered system model. In Section \ref{sec:results}, we present analytical and simulation results and their discussion. Next, we summarize the paper with main concluding remarks in Section \ref{sec:conclusion}. Finally, the analytical OP expressions over the RF/RF channels and of the asymptotic outage reduction are presented in Appendices \ref{App:1} and \ref{App:2}, respectively.

\begin{figure}[!t]
	\centering
	\includegraphics[width=1.0\columnwidth]{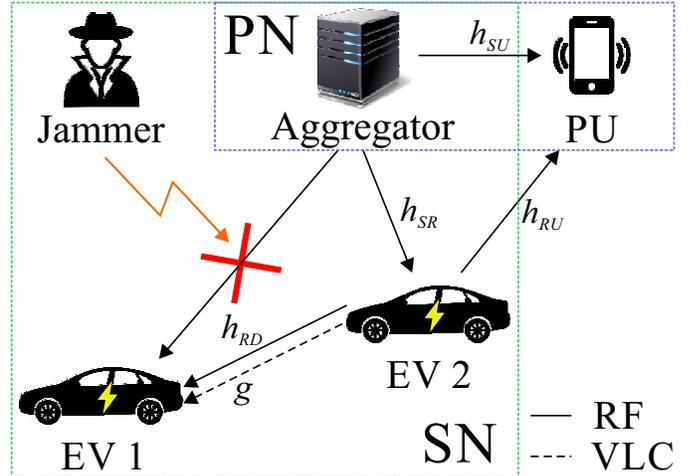}
	\caption{The system model of the outdoor CR-based electric vehicular network.}
	\label{fig:system}
\end{figure}

\section{System model}
\label{sec:system_model}
The adopted system model represents a downlink outdoor underlay cognitive radio network scenario represented by a mixture of the primary and secondary networks consisting of an aggregator ({\it i.e.}, a source node denoted by $S$), two electrical vehicles (EV 1 and EV 2), a jamming node ($J$) and a primary user ($U$). The network scenario can be described as follows. The primary network (PN) utilizes a given set of frequencies ({\it i.e.}, a pool of the frequencies) for communication purposes. As being part of the PN, the aggregator attempts to send information to the destination node ({\it i.e.}, EV 1 denoted by $D$) using one vacant frequency from the pool while the other EV ({\it i.e.}, EV 2) is idle ({\it e.g.}, parked and getting charged). At this moment, $J$ suppresses the communication link between $S$ and $D$. To overcome this, $S$ selects another frequency ({\it e.g.}, it can be vacant or occupied) from the pool to reach the jammed EV via the idle EV 2 that will act as a relay node (in the detect-and-forward mode and denoted by $R$). For the sake of practicality of the system setup, we consider four (4) cases given below:
\begin{itemize}
	\item {\it Case 1}: the newly selected frequencies are already being used in the PN such that the establishment of an end-to-end $S$-to-$D$ relay-assisted communication session causes interference to the primary nodes exploiting these frequencies. Thus, underlay CR mode is needed.
	\item {\it Case 2}: one of the selected frequencies is vacant in the PN such that $S$ can be assigned with it to establish the communication link between $S$ and $R$ with no interference at $U$. On the other hand, the other frequency utilized for the $R$-to-$D$ communication session is occupied and, thus, may lead to the intolerable level of interference at $U$. Then, the power constraint at $R$ is required.
	\item {\it Case 3}: this case is similar to {\it Case 2} with only difference, {\it i.e.}, the interference at $U$ is caused by $S$ only that needs to be power-restricted.
	\item {\it Case 4}: the frequencies selected from the pool for an end-to-end $S$-to-$D$ relay-assisted communication session are vacant and can be exploited for interference-free communication in the PN.
\end{itemize}
The end-to-end communication is realized over two identical time slots, with duration of $T/2$ each, where $T$ stands for the time period over which the channel estimates are assumed to remain constant. To avoid intolerable interference in the PN, the aggregator and EV 2, {\it i.e.}, $S$ and $R$, have to restrict their transmit power levels, if needed, as  
\begin{equation}
\label{relay_power}
P_i = \min\left(\bar{P}_i, \frac{K_i}{|h_{iU}|^2}\right),~i\in\{S,R\},
\end{equation}
where $K_i = I_{U}d_{iU}^{\tau}$, with $I_{U}$ standing for the interference temperature constraint (ITC) defined at $U$\footnote{Without loss of generality, we assume that orthogonal resources are assigned to all primary users. Hence, the SN communication session creates interference to only one primary receiver, {\it i.e.}, $U$.}, and $\bar{P}_i$ is the maximum allowed transmit power at node\footnote{This implies that the aggregator is also involved into the secondary network (SN) during the communication session operating on the chosen frequency channel. In other words, the aggregator can be treated as a gateway node between the primary and secondary networks.} $i$. $h_{iU}$, $d_{iU}$ and $\tau$ indicate the channel coefficient between $U$ and node $i$, the corresponding distance and the path-loss exponent, respectively. The communication sessions in the PN and between the aggregator and relay are organized in the RF spectrum due to its wide coverage and non-line-of-sight (NLOS) nature while the EVs communicate with each other through the RF or VLC channels.

The RF links are modeled by considering imperfect channel state information (CSI) as follows \cite{galymTVT2020}
\begin{align}
h_i = \tilde{h}_i + \epsilon,~\forall \in \mathcal{A} = \{SR,RD,SU,RU\},
\end{align}
where $\tilde{h}_i$ and $\epsilon$ denote the channel estimate and the estimation error, with $\mathcal{CN}\left(0, \sigma_{\epsilon}^2\right)$, respectively.

The received signal at $R$ ({\it i.e.}, EV 2) can be expressed as
\begin{align}
\label{yR}
y_R &= \sqrt{\frac{P_S}{d_{SR}^{\tau}}} h_{SR} x + n_R \nonumber\\
&= \sqrt{\frac{P_S}{d_{SR}^{\tau}}} \tilde{h}_{SR} x  + \sqrt{\frac{P_S}{d_{SR}^{\tau}}} \epsilon x + n_R,
\end{align}
where $h_{SR}$, $d_{SR}$, $x$ and $n_R$ denote the channel coefficient between $S$ and $R$, the corresponding  distance, the message dedicated to $D$ and the additive white Gaussian noise (AWGN) term at $R$, with variance $\sigma_R^2$, respectively. The corresponding signal-to-interference-plus-noise ratio (SINR) can be given by
\begin{align}
\label{gammaR}
\gamma_R &= \frac{P_S |\tilde{h}_{SR}|^2}{P_S \sigma_{\epsilon}^2 + d_{SR}^{\tau} \sigma_R^2}.
\end{align} 

It is reasonable to assume that $D$ ({\it i.e.}, EV 1) can acquire the RF and VLC CSI estimates\footnote{In different network scenarios, it is common for the nodes to exploit pilot signals in a cyclic manner for the channel estimation and synchronization purposes \cite[\S 12.3.1]{Jayaweera}. Thus, secondary nodes are also able to acquire their own channel estimates between the primary/secondary transmitters and themselves through participating in this training process \cite{sultanTVT}.} and then send the feedback to the relay which detects the received message using Eq. \eqref{gammaR} and transmits the data over the best channel\footnote{Since EVs are deployed with RF and VLC transceivers, selecting the best channel is identical to the best relay selection scheme \cite{best_relay}.}. Hence, the received RF signal at $D$ can be written as 
{\allowdisplaybreaks
\begin{align}
y_D^{\rm RF} &= \sqrt{\frac{P_R}{d_{RD}^{\tau}}} h_{RD} x + n_D \nonumber\\
&= \sqrt{\frac{P_R}{d_{RD}^{\tau}}} \tilde{h}_{RD} x 
+ \sqrt{\frac{P_R}{d_{RD}^{\tau}}} \epsilon x 
+ n_D,
\end{align}}where $h_{RD}$, $d_{RD}$ and $n_D$ denote the channel coefficient between $R$ and $D$, the corresponding distance, and the AWGN term, with variance $\sigma_D^2$, respectively. The corresponding SINR in the RF domain can be expressed as
\begin{align}
\label{gammaD}
\gamma_D^{\rm RF} &= \frac{P_R |\tilde{h}_{RD}|^2}{P_R \sigma_{\epsilon}^2 + d_{RD}^{\tau} \sigma_D^2}.
\end{align}

Regarding the VLC communication link, we consider outdoor vehicular visible light communications (V2LC), where a point-to-point communication link is established between a headlight of EV 2 and the tail of EV 1. The headlight uses light-emitting diode (LED) as the transmitter and a positive-intrinsic-negative photodetector (PIN PD) is installed on the tail as the receiver. Such a link experiences Rician fading as experimentally verified in \cite{RicianVLC}. Therefore, the received signal can be expressed as 
\begin{align}
y_D^{\rm VLC} &= \rho g_L g_R x + n_I \nonumber\\
&= \rho g_L \tilde{g}_R x + \rho g_L \epsilon x + n_I,
\end{align}
where $g_L$ is the path-loss following Lambertian emission, $\rho$ denotes the effective photo-current conversion ratio, $g_R$ is the atmospheric turbulence, and the $n_I$ is a Gaussian variable with variance $\sigma_{I}^2$. Therefore, the corresponding SNR is given by
\begin{align}
\gamma_D^{\rm VLC} = \frac{(\rho g_L \tilde{g}_R )^2 P_t}{(\rho g_L \epsilon )^2 P_t + \sigma_{I}^2},
\end{align}
where $P_t$ denotes the power of message $x$. The path-loss $g_L$ is modeled as \cite{PTL}
\begin{equation}
g_L = \begin{cases}
\begin{array}{ll}
\frac{(m+1) A_{\rm PD}}{2 \pi d_{RD}^2} 
\frac{\cos^m(\phi) T_s(\psi)}{\left( \cos(\psi) g(\psi) \right)^{-1}}, & {\rm if}~0\le \psi \le \Psi,\\
0, & {\rm if}~\psi \ge \Psi,
\end{array}
\end{cases}
\end{equation}
where $A_{\rm PD}$ is the PD detection area, $\phi$ is the angle of irradiance, $\psi$ is the angle of incidence, $T_s(\psi)$ is the optical filter gain, $g(\psi)$ is the concentrator gain, and $\Psi$ is the field-of-view (FOV).

The probability density function (PDF) of the Rician atmospheric turbulence $g_R$ is expressed as
\begin{align}
f_{g_R} \left( h_R \right) = \frac{h_R}{N} \exp\left(- \frac{h_R^2 + s^2}{2N}\right) I_0\left(\frac{h_R s}{N}\right),
\end{align}
where $I_0$ denotes the modified first kind Bessel function of order zero while $s$ and $N$ stand for the characteristic parameters of the Rician distribution. Moreover, we define another parameter, known as shape parameter $K = \frac{s^2}{2N}$, which represents the ratio of the power arrived from the line-of-sight (LOS) path to the power corresponding to the remaining NLOS component. Then, substituting $g_R = \frac{g}{g_L}$, the PDF of the overall channel coefficient can be written as
\begin{align}
f_{g} \left( g \right) &= f_{g} \left( \frac{g}{g_L} \right)\nonumber\\ 
&= \frac{g}{g_L N} \exp\left(- \frac{g^2 + g_L s^2}{2 g_L^2 N}\right) I_0\left(\frac{g s}{g_L N}\right).
\end{align}

The noise variance $\sigma_I^2$ of the PIN PD is the summation of the variances of the shot and thermal noises given as
{\allowdisplaybreaks
\begin{align}
\sigma_{\rm shot}^2 &= 2q \rho I_{bg} I_2 R_b,\\
\sigma_{\rm thermal}^2 &= \frac{8 \pi k T_k \eta A_{\rm PD} I_2 R_b^2}{G} \nonumber\\
&~~~+ \frac{16 \pi^2 k T_k \Gamma \eta^2 A_{\rm PD}^2 I_3 R_b^3}{g_m},
\end{align}}where $q$ is the electronic charge constant, $I_{bg}$ is the background current, $I_2$ and $I_3$ are the noise bandwidth factors, $R_b$ is the data rate, $k$ is the Boltzmann's constant, $T_k$ is the absolute temperature, $\eta$ is the fixed capacitance, $\Gamma$ is the field effect transistor (FET) channel noise factor, $G$ is the gain of open-loop voltage, and $g_m$ is the FET transconductance.

\section{Performance analysis}
\label{sec:analysis}
In this section, we derive analytical expression of the outage probability of {\it Case 1} since it involves both CR-based power constraints deployed at $S$ and $R$. The outage expressions for {\it Cases 2-4} can be derived by following the same approach.

Since the best channel is selected for transmission, the effective SNR can be expressed as
\begin{align}
\gamma^{\rm DF} &= \min\left(\gamma_R, \gamma_D \right),
\end{align}
where 
\begin{align}
\gamma_D &=\max\left(\gamma_D^{\rm RF}, \gamma_D^{\rm VLC}\right).
\end{align} 

The outage probability is given as
\begin{align}
\label{outage}
P_{\rm out} &= {\rm Pr}\left(\gamma^{DF} < v\right)\nonumber\\
&={\rm Pr}\left(\min\left(\gamma_R, \gamma_D \right) < v\right)\nonumber\\
&=1 - {\rm Pr}\left(\min\left(\gamma_R, \gamma_D \right) > v\right)\nonumber\\
&=1 - {\rm Pr}\left(\gamma_R > v\right) 
{\rm Pr}\left(\gamma_D > v\right) \nonumber\\
&=1 - {\rm Pr}\left(\gamma_R > v\right) 
{\rm Pr}\left(\max\left(\gamma_D^{\rm RF}, \gamma_D^{\rm VLC}\right) > v\right) \nonumber\\
&=1 - {\rm Pr}\left(\gamma_R > v\right) 
\left[
1 - {\rm Pr}\left(\max\left(\gamma_D^{\rm RF}, \gamma_D^{\rm VLC}\right) < v\right)
\right] \nonumber\\
&=1 - {\rm Pr}\left(\gamma_R > v\right) 
\left[
1 - {\rm Pr}\left(\gamma_D^{\rm RF} < v\right)
{\rm Pr}\left(\gamma_D^{\rm VLC} < v\right)
\right] \nonumber\\
&=1 - 
\left[
1- 
\underset{A}{\underbrace{{\rm Pr}\left(\gamma_R < v\right)}}
\right] \nonumber\\
&
\hspace{1.3cm}
\times \left[
1 - \underset{B_1}{\underbrace{{\rm Pr}\left(\gamma_D^{\rm RF} < v\right)}}
\underset{B_2}{\underbrace{{\rm Pr}\left(\gamma_D^{\rm VLC} < v\right)}}
\right],
\end{align}
where $\mathcal{R}_{\rm th}$ and $v = 2^{2\mathcal{R}_{\rm th}} - 1$ denote the data rate threshold and its corresponding SNR-based value.

The term $A$ can be expressed as
\begin{align}
\label{term_A}
A &= {\rm Pr}\left(\frac{P_S |\tilde{h}_{SR}|^2}{P_S \sigma_{\epsilon}^2 + d_{SR}^{\tau} \sigma_R^2} < v \right) = A_1 + A_2,
\end{align}
where the term $A_1$ is calculated as
\begin{align}
\label{A1}
A_1 &= {\rm Pr}\left(\frac{P_S |\tilde{h}_{SR}|^2}{P_S \sigma_{\epsilon}^2 + d_{SR}^{\tau} \sigma_R^2} < v, P_S < \frac{K_S }{|h_{SU}|^2} \right) \nonumber\\
&= {\rm Pr}\left(|\tilde{h}_{SR}|^2 < \frac{v \left(P_S \sigma_{\epsilon}^2 + d_{SR}^{\tau} \sigma_R^2\right)}{P_S}, |h_{SU}|^2 < \frac{K_S}{P_S} \right)\nonumber\\
&= \left[
1 - \exp\left(- \tilde{\lambda}_{SR} v \left( \sigma_{\epsilon}^2 + \frac{d_{SR}^{\tau} \sigma_R^2}{P_S}\right) \right)
\right] \nonumber\\
&~~~ \times 
\left[
1 - \exp\left(- \lambda_{SU} \frac{K_S}{P_S}\right)
\right],
\end{align}
and the term $A_2$ is given by 
\begin{align}
\label{A2}
A_2 &= {\rm Pr}\left(\frac{K_S |\tilde{h}_{SR}|^2}{K_S \sigma_{\epsilon}^2 + d_{SR}^{\tau} \sigma_R^2 |h_{SU}|^2} < v, P_S > \frac{K_S}{|h_{SU}|^2} \right) \nonumber\\
&= {\rm Pr}\left(|\tilde{h}_{SR}|^2 < v \left(\sigma_{\epsilon}^2 + \frac{d_{SR}^{\tau} \sigma_R^2 }{K_S} |h_{SU}|^2 \right), \right. \nonumber\\
&\hspace{3cm} \left. |h_{SU}|^2 > \frac{K_S}{P_S} \right)\nonumber\\
&= \exp\left(-\lambda_{SU} \frac{K_S}{P_S} \right) 
- 
\Sigma_S  
\exp\left(-\lambda_{SU} \frac{K_S}{P_S} \right)
\nonumber\\
&~~~ \times
\exp\left(- \tilde{\lambda}_{SR} v \left(
\sigma_{\epsilon}^2 + \frac{d_{SR}^{\tau} \sigma_R^2 }{P_S}
\right) \right),
\end{align}
where $\Sigma_S = \left(1 + 
\frac{\tilde{\lambda}_{SR} v  d_{SR}^{\tau} \sigma_R^2 }{\lambda_{SU} K_S}
\right)^{-1}$. Note that $|h_i|^2, \forall i \in \mathcal{A},$ follows an exponential probability distribution \cite{sultanTVT}, and ${\lambda}_i$ is the rate parameter of the random variable (RV) $|{h}_i|^2$. 

Combining \eqref{A1} and \eqref{A2} together, $A$ can be written as 
\begin{multline}
	\label{A_final}
	A = 1 - \exp\left(- \tilde{\lambda}_{SR} v \left(\sigma_{\epsilon}^2 + \frac{d_{SR}^{\tau} \sigma_R^2 }{P_S} \right) \right) \\
	\left[1 - \left(
	1 - \Sigma_S
	\right) \exp\left( - \lambda_{SU} \frac{K_S}{P_S} \right)  \right].
\end{multline}

Similar to the term $A$ in \eqref{term_A}, the cumulative density function (CDF) $B_1$ of $\gamma_D^{\rm RF}$ can be computed as \cite{sultanCOMML1}
\begin{align}
B_1 = {\rm Pr}\left(\gamma_D^{\rm RF} < v \right) = B_{11} + B_{12},
\end{align}
where, due to the independence of the RVs in $B_{11}$, it can be simply expressed as
\begin{align}
B_{11} &= {\rm Pr}\left(\frac{P_R |\tilde{h}_{RD}|^2}{P_R \sigma_{\epsilon}^2 + d_{RD}^{\tau} \sigma_{D}^2} < v, P_R < \frac{K_R}{|h_{RU}|^2}\right) \nonumber\\
&= {\rm Pr}\left(|\tilde{h}_{RD}|^2 < v \sigma_{\epsilon}^2 + 
\frac{d_{RD}^{\tau} \sigma_{D}^2}{P_R}, |h_{RU}|^2 < \frac{K_R}{P_R}\right)\nonumber\\
&=\left[
1 - \exp\left( - \tilde{\lambda}_{RD} v \left( \sigma_{\epsilon}^2 + \frac{ d_{RD}^{\tau} \sigma_{D}^2}{P_R} \right) \right)
\right]
\nonumber\\
&~~~
\times\left[1 - \exp\left( - \lambda_{RU} \frac{K_R}{P_R}\right)\right].
\end{align}
On the other hand, after some algebraic manipulations, the term $B_{12}$ can be computed as
\begin{align}
B_{12} &= {\rm Pr}\left(\frac{K_R |\tilde{h}_{RD}|^2}{K_R \sigma_{\epsilon}^2 + d_{RD}^{\tau} \sigma_{D}^2 |h_{RU}|^2} < v, P_R > \frac{K_R}{|h_{RU}|^2}\right) \nonumber\\
&= {\rm Pr}\left(|\tilde{h}_{RD}|^2 < v \left(\sigma_{\epsilon}^2 + \frac{ d_{RD}^{\tau} \sigma_{D}^2 }{K_R} |h_{RU}|^2 \right), \right. \nonumber\\
&\hspace{4.5cm}\left. |h_{RU}|^2 > \frac{K_R}{P_R}\right), \nonumber\\
&= \exp\left( - \lambda_{RU} \frac{K_R}{P_R}\right) 
- 
\Sigma_R
\exp\left( - \lambda_{RU} \frac{K_R}{P_R}\right) \nonumber\\
&~~~\times 
\exp\left(- \tilde{\lambda}_{RD} v \left(\sigma_{\epsilon}^2 + \frac{ d_{RD}^{\tau} \sigma_{D}^2 }{P_R}\right) \right),
\end{align}
where $\Sigma_R = \left(1 + \frac{\tilde{\lambda}_{RD} v d_{RD}^{\tau} \sigma_{D}^2 }{\lambda_{RU} K_R} \right)^{-1}$.
 
Finally, the CDF of $\gamma_D^{\rm RF}$ can be expressed as
\begin{multline}
\label{B1}
B_{1} = 1 - \exp\left(- \tilde{\lambda}_{RD} v \left(\sigma_{\epsilon}^2 + \frac{ d_{RD}^{\tau} \sigma_{D}^2}{P_R} \right) \right)
\\
\times 
\left[
1 - \left(1 - \Sigma_R
\right) \exp\left(- \lambda_{RU} \frac{K_R}{P_R}\right)
\right].
\end{multline}

The CDF $B_2$ of the RV $\gamma_D^{\rm VLC}$ can be calculated as
\begin{align}
B_2 &= {\rm Pr}\left(\frac{(\rho g_L \tilde{g}_R )^2 P_t}{(\rho g_L \epsilon )^2 P_t + \sigma_{I}^2} < v\right) \nonumber\\
&= {\rm Pr}\left( g_R < \frac{\sqrt{v \left((\rho g_L \sigma_{\epsilon} )^2 P_t + \sigma_{I}^2\right)} }{\rho g_L \sqrt{P_t}}\right) \nonumber\\
&= 1 - Q_1\left(\frac{s}{\sqrt{N}}, \frac{\sqrt{v \left((\rho g_L \sigma_{\epsilon} )^2 P_t + \sigma_{I}^2\right)} }{\sqrt{N P_t} \rho g_L}\right),
\end{align}
where $Q_1$ is the Marcum Q-function of order 1 \cite{marcum}.

Then, substituting the terms $A$, $B_1$ and $B_2$ into Eq. \eqref{outage}, the outage probability for {\it Case 1} can be expressed in closed form as in Eq. \eqref{outage1_Case1}, shown at the top of the next page. The outage expressions for {\it Cases 2-4} are presented in Eqs. \eqref{outage1_Case2}-\eqref{outage1_Case4}, shown at the top of the next page.

\begin{figure*}[!t]
	\begin{align}
	\label{outage1_Case1}
	P^{{\it Case~}{\rm 1}}_{\rm out}(v) &= 1 -  
	\exp\left(- \tilde{\lambda}_{SR} v \left(\sigma_{\epsilon}^2 + \frac{d_{SR}^{\tau} \sigma_R^2 }{P_S} \right) \right) 
	\left[1 - \left(
	1 - \Sigma_S
	\right) \exp\left( - \lambda_{SU} \frac{K_S}{P_S} \right)  \right]
	 \nonumber \\
	&~~~~~~~~\times\left\lbrace   
	1 -
	\left[
	1 - Q_1\left(\frac{s}{\sqrt{N}}, \frac{\sqrt{v \left((\rho g_L \sigma_{\epsilon} )^2 P_t + \sigma_{I}^2\right)} }{\sqrt{N P_t} \rho g_L}\right)
	\right] \right. \nonumber\\
	&\hspace{3.5cm}\times \left.
	\left[
	1 - \exp\left(- \tilde{\lambda}_{RD} v \left(\sigma_{\epsilon}^2 + \frac{ d_{RD}^{\tau} \sigma_{D}^2}{P_R} \right) \right)
	\left[
	1 - \left(1 - \Sigma_R
	\right) \exp\left(- \lambda_{RU} \frac{K_R}{P_R}\right)
	\right]
	\right]
	\right\rbrace 
	\end{align}
	\hrulefill
	\begin{multline}
	\label{outage1_Case2}
	P^{{\it Case~}{\rm 2}}_{\rm out}(v) = 1 - 
	\exp\left(- \tilde{\lambda}_{SR} v \left( \sigma_{\epsilon}^2 + \frac{d_{SR}^{\tau} \sigma_R^2}{P_S}\right) \right)
	\left\lbrace   
	1 -
	\left[
	1 - Q_1\left(\frac{s}{\sqrt{N}}, \frac{\sqrt{v \left((\rho g_L \sigma_{\epsilon} )^2 P_t + \sigma_{I}^2\right)} }{\sqrt{N P_t} \rho g_L}\right)
	\right] \right. \\ 
	\left.
	\times
	\left[
	1 - \exp\left(- \tilde{\lambda}_{RD} v \left(\sigma_{\epsilon}^2 + \frac{ d_{RD}^{\tau} \sigma_{D}^2}{P_R} \right) \right)
	\left[
	1 - \left(1 - \Sigma_R
	\right) \exp\left(- \lambda_{RU} \frac{K_R}{P_R}\right)
	\right]
	\right]
	\right\rbrace 
	\end{multline}
	\hrulefill
	\begin{multline}
	\label{outage1_Case3}
	P^{{\it Case~}{\rm 3}}_{\rm out}(v) = 1 - \left\lbrace  
	\exp\left(- \tilde{\lambda}_{SR} v \left(\sigma_{\epsilon}^2 + \frac{d_{SR}^{\tau} \sigma_R^2 }{P_S} \right) \right) 
	\left[1 - \left(
	1 - \Sigma_S
	\right) \exp\left( - \lambda_{SU} \frac{K_S}{P_S} \right)  \right]
	\right\rbrace  \\
	\times\left\lbrace   
	1 -
	\left[
	1 - Q_1\left(\frac{s}{\sqrt{N}}, \frac{\sqrt{v \left((\rho g_L \sigma_{\epsilon} )^2 P_t + \sigma_{I}^2\right)} }{\sqrt{N P_t} \rho g_L}\right)
	\right] 
	\left[
	1 - \exp\left( - \tilde{\lambda}_{RD} v \left( \sigma_{\epsilon}^2 + \frac{ d_{RD}^{\tau} \sigma_{D}^2}{P_R} \right) \right)
	\right]
	\right\rbrace 
	\end{multline}
	\hrulefill
	\begin{multline}
	\label{outage1_Case4}
	P^{{\it Case~}{\rm 4}}_{\rm out}(v) = 1 - 
	\exp\left(- \tilde{\lambda}_{SR} v \left( \sigma_{\epsilon}^2 + \frac{d_{SR}^{\tau} \sigma_R^2}{P_S}\right) \right)
	\\
	\times\left\lbrace   
	1 -
	\left[
	1 - Q_1\left(\frac{s}{\sqrt{N}}, \frac{\sqrt{v \left((\rho g_L \sigma_{\epsilon} )^2 P_t + \sigma_{I}^2\right)} }{\sqrt{N P_t} \rho g_L}\right)
	\right] 
	\left[
	1 - \exp\left(- \tilde{\lambda}_{RD} v \left(\sigma_{\epsilon}^2 + \frac{ d_{RD}^{\tau} \sigma_{D}^2}{P_R} \right) \right)
	\right]
	\right\rbrace 
	\end{multline}
	\hrulefill
\end{figure*}

Now, using Eq. \eqref{outage1_Case1}, the average throughput can be evaluated as
\begin{align}
	\label{throuput}
	\mathcal{C} &= \mathcal{R} \left( 1 - \mathcal{P}_{\rm out}\right),
\end{align}
where $\mathcal{R}$ denotes the achievable data rate given by
\begin{equation}
\mathcal{R} = \frac{1}{2} \log_2 \left(1 + \min\left(\gamma_R, \max\left(\gamma_D^{\rm RF},\gamma_D^{\rm VLC}\right)\right)\right),
\end{equation} 
where the factor $\frac{1}{2}$ implies that the end-to-end transmission requires two time slots.

\subsection{Asymptotic Analysis}
To obtain meaningful insights on the impact of the involved fading parameters on the network performance, we investigate the system performance in the high SNR regime. By setting $\{P_S, P_R, P_t\} \to \infty$, the high SNR approximation of the end-to-end OP for all {\it Cases} can be expressed as the $(a)$ expressions given by Eqs. \eqref{outageY_Case1}, \eqref{outageY_Case2}, \eqref{outageY_Case3} and \eqref{outageY_Case4}, where the terms $\Sigma_i,~i\in\{S,R\},$ in the OP expressions represent the impact of CR communication\footnote{Note that {\it Case~4} presumes that the secondary $S$-to-$D$ communication does not cause any interference in the PN.}. Moreover, further approximation given by the $(b)$ expressions in Eqs. \eqref{outageY1_Case1}, \eqref{outageY1_Case2} and \eqref{outageY1_Case3} represent the scenario when the transmit power levels $P_S$ and $P_R$ are not constrained by Eq. \eqref{relay_power}, {\it i.e.}, the secondary communication does not harm the primary receiver ($I_{U} \to \infty$). Another useful insight is that the $(b)$ expressions related to all {\it Cases} coincide with each other, and this implies that they imitate {\it Case 4}, as expected. As it can be seen from the general expressions of the approximated OP, it depends on the quality of CSI estimates and the maximum level of interference that $U$ can tolerate. {Moreover, the outage expressions can be further simplified by applying $e^{x} = 1 + x$ for small $x$ as in Eqs. \eqref{outageY2_Case1}, \eqref{outageY2_Case2}, \eqref{outageY2_Case3} and \eqref{outageY2_Case4}, shown at the top of the next page,} where $\tilde{\lambda}_i v \sigma_{\epsilon}^2 \approx 0, \forall i=\{SR,RD\}$. Finally, in the case of perfect CSI, network failure will not occur, {\it i.e.}, $P_{\rm out} \to 0$ since $Q_1(a,0) = 1$ for any $a$ due to $\int_{0}^{\infty} f(x) {\rm d}x = 1$.

\begin{figure*}[!t]
	\begin{align}
	\label{outageY_Case1}
	P^{{\it Case~}{\rm 1:~asymp}}_{\rm out} (v) 
	&
	\underset{\{P_S, P_R, P_t\} \to \infty}{\stackrel{(a)}{=}} 
	1 -   
	\exp\left(- \tilde{\lambda}_{SR} v \sigma_{\epsilon}^2 \right) 
	\left[1 - \left(
	1 - \Sigma_S
	\right) \exp\left( - \lambda_{SU} \frac{K_S}{P_S} \right)  \right]
	\nonumber \\
	& 
	\times \left\lbrace   
	1 -
	\left[
	1 - Q_1\left(\frac{s}{\sqrt{N}}, \sqrt{\frac{v \sigma_{\epsilon}^2 }{N }} \right)
	\right] 
	\left[
	1 - \exp\left(- \tilde{\lambda}_{RD} v \sigma_{\epsilon}^2  \right)
	\left[
	1 - \left(1 - \Sigma_R
	\right) \exp\left(- \lambda_{RU} \frac{K_R}{P_R}\right)
	\right]
	\right]
	\right\rbrace 
	\\
	\label{outageY1_Case1}
	& \underset{{\rm no}~I_{U} }{\stackrel{(b)}{=}}
	1 -   
	\exp\left(- \tilde{\lambda}_{SR} v \sigma_{\epsilon}^2 \right) 
	\left\lbrace   
	1 -
	\left[
	1 - Q_1\left(\frac{s}{\sqrt{N}}, \sqrt{\frac{v \sigma_{\epsilon}^2 }{N }} \right)
	\right] 
	\left[
	1 - \exp\left(- \tilde{\lambda}_{RD} v \sigma_{\epsilon}^2  \right)
	\right]
	\right\rbrace  
	\\
	\label{outageY2_Case1}
	& \underset{\sigma_{\epsilon}^2 \to 0}{\stackrel{(c)}{=}}
	1 -   
	\left(1 - \tilde{\lambda}_{SR} v \sigma_{\epsilon}^2 \right) 
	\left\lbrace   
	1 -
	\tilde{\lambda}_{RD} v \sigma_{\epsilon}^2
	\left[
	1 - Q_1\left(\frac{s}{\sqrt{N}}, \sqrt{\frac{v \sigma_{\epsilon}^2 }{N }} \right)
	\right] 
	\right\rbrace \to 0
	\end{align}
	\hrulefill
	\begin{align}
	\label{outageY_Case2}
	P^{{\it Case~}{\rm 2:~asymp}}_{\rm out} (v) 
	\underset{\{P_S, P_R, P_t\} \to \infty}{\stackrel{(a)}{=}} &
	1 -   
	\exp\left(- \tilde{\lambda}_{SR} v \sigma_{\epsilon}^2 \right) \left\lbrace   
	1 -
	\left[
	1 - Q_1\left(\frac{s}{\sqrt{N}}, \sqrt{\frac{v \sigma_{\epsilon}^2 }{N }} \right)
	\right] \right.
	\nonumber \\
	& \hspace{3.cm} \left. \times 
	\left[
	1 - \exp\left(- \tilde{\lambda}_{RD} v \sigma_{\epsilon}^2  \right)
	\left[
	1 - \left(1 - \Sigma_R
	\right) \exp\left(- \lambda_{RU} \frac{K_R}{P_R}\right)
	\right]
	\right]
	\right\rbrace 
	\\
	\label{outageY1_Case2}
	\underset{{\rm no}~I_{U} }{\stackrel{(b)}{=}}
	&
	1 -   
	\exp\left(- \tilde{\lambda}_{SR} v \sigma_{\epsilon}^2 \right) 
	\left\lbrace   
	1 -
	\left[
	1 - Q_1\left(\frac{s}{\sqrt{N}}, \sqrt{\frac{v \sigma_{\epsilon}^2 }{N }} \right)
	\right] 
	\left[
	1 - \exp\left(- \tilde{\lambda}_{RD} v \sigma_{\epsilon}^2  \right)
	\right]
	\right\rbrace  
	\\
	\label{outageY2_Case2}
	\underset{\sigma_{\epsilon}^2 \to 0}{\stackrel{(c)}{=}}
	&
	1 -   
	\left(1 - \tilde{\lambda}_{SR} v \sigma_{\epsilon}^2 \right) 
	\left\lbrace   
	1 -
	\tilde{\lambda}_{RD} v \sigma_{\epsilon}^2
	\left[
	1 - Q_1\left(\frac{s}{\sqrt{N}}, \sqrt{\frac{v \sigma_{\epsilon}^2 }{N }} \right)
	\right] 
	\right\rbrace \to 0
	\end{align}
	\hrulefill
\end{figure*}

\begin{figure*}[!t]
	\begin{align}
	\label{outageY_Case3}
	P^{{\it Case~}{\rm 3:~asymp}}_{\rm out} (v) 
	\underset{\{P_S, P_R, P_t\} \to \infty}{\stackrel{(a)}{=}} 
	& 
	1 -   
	\exp\left(- \tilde{\lambda}_{SR} v \sigma_{\epsilon}^2 \right) 
	\left[1 - \left(
	1 - \Sigma_S
	\right) \exp\left( - \lambda_{SU} \frac{K_S}{P_S} \right)  \right]
	\nonumber \\
	& \times \left\lbrace   
	1 -
	\left[
	1 - Q_1\left(\frac{s}{\sqrt{N}}, \sqrt{\frac{v \sigma_{\epsilon}^2 }{N }} \right)
	\right] 
	\left[
	1 - \exp\left(- \tilde{\lambda}_{RD} v \sigma_{\epsilon}^2  \right)
	\right]
	\right\rbrace 
	\\
	\label{outageY1_Case3}
	\underset{{\rm no}~I_{U} }{\stackrel{(b)}{=}}
	& 1 -   
	\exp\left(- \tilde{\lambda}_{SR} v \sigma_{\epsilon}^2 \right) 
	\left\lbrace   
	1 -
	\left[
	1 - Q_1\left(\frac{s}{\sqrt{N}}, \sqrt{\frac{v \sigma_{\epsilon}^2 }{N }} \right)
	\right] 
	\left[
	1 - \exp\left(- \tilde{\lambda}_{RD} v \sigma_{\epsilon}^2  \right)
	\right]
	\right\rbrace  
	\\
	\label{outageY2_Case3}
	\underset{\sigma_{\epsilon}^2 \to 0}{\stackrel{(c)}{=}}
	& 
	1 -   
	\left(1 - \tilde{\lambda}_{SR} v \sigma_{\epsilon}^2 \right) 
	\left\lbrace   
	1 -
	\tilde{\lambda}_{RD} v \sigma_{\epsilon}^2
	\left[
	1 - Q_1\left(\frac{s}{\sqrt{N}}, \sqrt{\frac{v \sigma_{\epsilon}^2 }{N }} \right)
	\right] 
	\right\rbrace \to 0
	\end{align}
	\hrulefill
	\begin{align}
	\label{outageY_Case4}
	P^{{\it Case~}{\rm 4:~asymp}}_{\rm out} (v) 
	\underset{\{P_S, P_R, P_t\} \to \infty}{\stackrel{(a)~{\rm and}~(b)}{=}} 
	& 
	1 -   
	\exp\left(- \tilde{\lambda}_{SR} v \sigma_{\epsilon}^2 \right) 
	\left\lbrace   
	1 -
	\left[
	1 - Q_1\left(\frac{s}{\sqrt{N}}, \sqrt{\frac{v \sigma_{\epsilon}^2 }{N }} \right)
	\right] 
	\left[
	1 - \exp\left(- \tilde{\lambda}_{RD} v \sigma_{\epsilon}^2  \right)
	\right]
	\right\rbrace 
	\\
	\label{outageY2_Case4}
	\underset{\sigma_{\epsilon}^2 \to 0}{\stackrel{(c)}{=}}
	&
	1 -   
	\left(1 - \tilde{\lambda}_{SR} v \sigma_{\epsilon}^2 \right) 
	\left\lbrace   
	1 -
	\tilde{\lambda}_{RD} v \sigma_{\epsilon}^2
	\left[
	1 - Q_1\left(\frac{s}{\sqrt{N}}, \sqrt{\frac{v \sigma_{\epsilon}^2 }{N }} \right)
	\right] 
	\right\rbrace \to 0
	\end{align}
	\hrulefill
\end{figure*}


\begin{table}[]
	\centering
	\caption{Simulation parameters.}
	\label{Table1}
	\begin{tabular}{|l|r|}
		\hline 
		Parameter & Value \\ 
		\hline \hline
		$S$-to-$R$ distance, $d_{SR}$ & 15 m\\
		\hline
		$S$-to-$U$ distance, $d_{SU}$ & 20 m\\
		\hline
		$R$-to-$D$ distance, $d_{RD}$ & $\{10;20\}$ m\\
		\hline
		$R$-to-$U$ distance, $d_{RU}$ & 5 m\\
		\hline
		Path-loss exponent, $\tau$ & 2.7\\
		\hline
		Shape index of the Rician distribution, $K$ & $\{0;5;10\}$ dB \\
		\hline
		Lambertian order, $m$ & 1 \\
		\hline
		LED semi-angle, $\Phi_{1/2}$ & 60$^\circ$\\
		\hline
		Optical concentrator gain, $g(\psi_k)$ & 1 \\ 
		\hline
		Optical filter gain, $T(\psi_k)$ & 1 \\ 
		\hline
		Effective photo-current conversion ratio, $\rho$ & 0.53 A/W \\ 
		\hline
		PIN PD detection area, $A_{\rm PD}$ & $10^{-4}$ m$^2$ \\
		\hline
		PIN PD FOV, $\Psi_{\textrm{FOV}}$ & 60$^\circ$ \\
		\hline
		Background current, $I_{bg}$ & $10^3$ $\mu$A\\
		\hline
		Noise bandwidth factor, $I_{2}$ & $0.562$\\
		\hline
		Noise bandwidth factor, $I_{3}$ & $0.0868$\\
		\hline
		Data rate, $R_{b}$ & $200$ Mbits/s\\
		\hline
		Absolute temperature, $T_{K}$ & 300 K\\
		\hline
		Fixed capacitance, $\eta$ & $112 \times 10^{-8}$ \\
		\hline
		FET channel noise factor, $\Gamma$ & $1.5$ \\
		\hline
		Gain of open-loop voltage, $G$ & $10$ \\
		\hline
		FET transconductance, $g_m$ & $30$ mS \\
		\hline
	\end{tabular}
\end{table}

\subsection{Reduction of Outage}
It is interesting to quantify the performance advantage of exploiting such a dual-hop CR network comprising two complementary wireless technologies. This advantage can be effectively evaluated in terms of communication reliability, {\it i.e.}, reduction of outage. With this in mind, we define the OP achievable by the dual-hop RF-based CR network as in Eqs. \eqref{rf-rf_Case1}-\eqref{rf-rf_Case4} (see Appendix \ref{App:1}). Therefore, the outage reduction can be defined as 
$$\Delta_{\rm out} = \left| P_{\rm out} - P_{\rm out}^{\rm RF/RF} \right|,$$ 
which is written in detail in Eqs. \eqref{reduction_Case1}-\eqref{reduction_Case4}, shown at the top of the next pages. Moreover, their high SNR approximation can be further expressed as in Eqs. \eqref{reduction_Case1_asymp}-\eqref{reduction_Case2_asymp} (for more detail, please refer to Appendix \ref{App:2}). From them, we can observe that the outage performance of the considered system model improves as both $\Sigma_S$ and $\Sigma_R$ tend to 1. Finally, it is worth noting that, unlike the other involved terms, the ITC term is the most critical one since the outage benefit severely decreases as $I_{U}$ increases and consequently results in performance saturation. 

\begin{figure*}
	\begin{multline}
	\label{reduction_Case1}
	\Delta_{\rm out}^{Case~{\rm 1}}(v) = 
	\left|
	-
	\exp\left(- \tilde{\lambda}_{SR} v \left(\sigma_{\epsilon}^2 + \frac{d_{SR}^{\tau} \sigma_R^2 }{P_S} \right) \right) \left[1 - \left(
	1 - \Sigma_S
	\right) \exp\left( - \lambda_{SU} \frac{K_S}{P_S} \right)  \right] 
	Q_1\left(\frac{s}{\sqrt{N}}, \frac{\sqrt{v \left((\rho g_L \sigma_{\epsilon} )^2 P_t + \sigma_{I}^2\right)} }{\sqrt{N P_t} \rho g_L}\right)
	\right. \\
	\left. \times 
	\left[ 
	1-
	\exp\left(- \tilde{\lambda}_{RD} v \left(\sigma_{\epsilon}^2 + \frac{ d_{RD}^{\tau} \sigma_{D}^2}{P_R} \right) \right) 
	\left(
	1 - \left(1 - \Sigma_R
	\right) \exp\left(- \lambda_{RU} \frac{K_R}{P_R}\right)
	\right)  
	\right]
	\right|
	\end{multline}
	\hrulefill
	\begin{multline}
	\label{reduction_Case2}
	\Delta_{\rm out}^{Case~{\rm 2}}(v) = 
	\left|
	-
	\exp\left(- \tilde{\lambda}_{SR} v \left(\sigma_{\epsilon}^2 + \frac{d_{SR}^{\tau} \sigma_R^2 }{P_S} \right) \right) 
	Q_1\left(\frac{s}{\sqrt{N}}, \frac{\sqrt{v \left((\rho g_L \sigma_{\epsilon} )^2 P_t + \sigma_{I}^2\right)} }{\sqrt{N P_t} \rho g_L}\right)
	\right. \\
	\left. \times 
	\left[ 
	1-
	\exp\left(- \tilde{\lambda}_{RD} v \left(\sigma_{\epsilon}^2 + \frac{ d_{RD}^{\tau} \sigma_{D}^2}{P_R} \right) \right) 
	\left(
	1 - \left(1 - \Sigma_R
	\right) \exp\left(- \lambda_{RU} \frac{K_R}{P_R}\right)
	\right)  
	\right]
	\right|
	\end{multline}
	\hrulefill
	\begin{multline}
	\label{reduction_Case3}
	\Delta_{\rm out}^{Case~{\rm 3}}(v) = 
	\left|
	-
	\exp\left(- \tilde{\lambda}_{SR} v \left(\sigma_{\epsilon}^2 + \frac{d_{SR}^{\tau} \sigma_R^2 }{P_S} \right) \right) \left[1 - \left(
	1 - \Sigma_S
	\right) \exp\left( - \lambda_{SU} \frac{K_S}{P_S} \right)  \right] 
	\right. \\
	\left. \times 
	Q_1\left(\frac{s}{\sqrt{N}}, \frac{\sqrt{v \left((\rho g_L \sigma_{\epsilon} )^2 P_t + \sigma_{I}^2\right)} }{\sqrt{N P_t} \rho g_L}\right)
	\left[ 
	1-
	\exp\left(- \tilde{\lambda}_{RD} v \left(\sigma_{\epsilon}^2 + \frac{ d_{RD}^{\tau} \sigma_{D}^2}{P_R} \right) \right)  
	\right]
	\right|
	\end{multline}
	\hrulefill
	\begin{multline}
	\label{reduction_Case4}
	\Delta_{\rm out}^{Case~{\rm 4}}(v) = 
	\left|
	-
	\exp\left(- \tilde{\lambda}_{SR} v \left(\sigma_{\epsilon}^2 + \frac{d_{SR}^{\tau} \sigma_R^2 }{P_S} \right) \right) 
	Q_1\left(\frac{s}{\sqrt{N}}, \frac{\sqrt{v \left((\rho g_L \sigma_{\epsilon} )^2 P_t + \sigma_{I}^2\right)} }{\sqrt{N P_t} \rho g_L}\right)
	\right. \\
	\left. \times 
	\left[ 
	1-
	\exp\left(- \tilde{\lambda}_{RD} v \left(\sigma_{\epsilon}^2 + \frac{ d_{RD}^{\tau} \sigma_{D}^2}{P_R} \right) \right)   
	\right]
	\right|
	\end{multline}
	\hrulefill
\end{figure*}

\section{Results Discussion}
\label{sec:results}
In this section, we present numerical results on the outage performance of the outdoor cognitive electric vehicular network. The simulation parameters are presented in Table \ref{Table1}. Without loss of generality, it is reasonable to assume that the LED and PIN PD comprising the secondary VLC communication link are aligned along a line perpendicular to the LED and PIN PD planes. Otherwise, the VLC channel gain will be decreased. Therefore, for simulation purposes, we assume two different distances, {\it i.e.}, 10 m and 20 m, with the corresponding VLC channel gains of -65 dB and -71 dB, respectively (calculated with respect to the simulation parameters given in Table \ref{Table1}). We also assume that, in the case of the RF channel, the transmit power levels at the source and relay nodes are the same, {\it i.e.}, $P = P_S = P_R$.

\begin{figure}[!t]
	\centering
	\includegraphics[width=1.0\columnwidth]{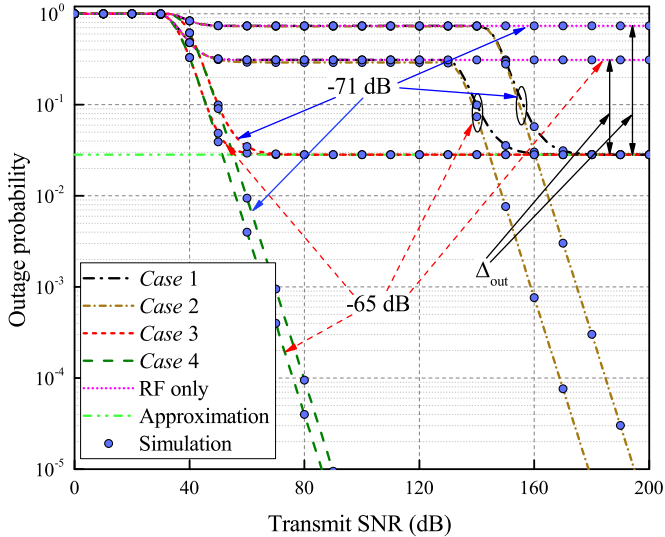}
	\caption{The outage probability versus the transmit SNR under perfect CSI for all {\it Cases} and different values of the VLC channel gain $h_L=\{-65;-71\}$ dB when $v = 2$ and $I_U = 15$ dB.}
	\label{res:OP1}
\end{figure}

In Fig. \ref{res:OP1}, we present some analytical and simulation results on the outage performance achievable by the proposed technique and conventional dual-hop RF-based scheme. The latter one is chosen as a benchmark to demonstrate the effectiveness of the proposed scheme. As it can be seen in Fig. \ref{res:OP1}, the mixed RF/VLC technique always outperforms the RF one for various CSI scenarios. We evaluate the OP metric versus the transmit SNR which is defined as $\xi^{\rm Rayl} = P/\sigma^2$ and $\xi^{\rm Rice} = P_t/\sigma_{I}^2$ (for simplicity, we assume the same transmit SNR values to both RF and VLC communication links, {\it i.e.}, $\xi = \xi^{\rm Rayl} = \xi^{\rm Rice}$)\footnote{With our assumptions and system parameters, it is reasonable to consider such high SNR values, {\it e.g.}, 200 dB, which can be easily obtained due to the corresponding small variances of the noise terms.}. First, we consider the outage performance when $d_{RD}=10$ m and $I_U = 15$ dB imposed at both $S$ and $R$.

In {\it Case~1}, the outage curve (a black dash-dot line) experiences two-stage saturation at 40 dB and 160 dB which can be explained as follows. The first saturation occurs due to the power restriction imposed at $R$ and, then, after 130 dB, the outage improves due to the availability of the VLC link, {\it i.e.}, the VLC-based SNR starts prevailing over the RF-based SNR. Such long saturation region can be explained by extremely small channel coefficients related to the VLC channel modeling. Moreover, since the VLC communication link does not interfere with the primary network, the OP performance starts improving by approaching the ideal case ({\it i.e.}, becomes parallel to the line in {\it Case 1} given by a green dash line) under the assumption that there is no ITC declared in the network. The second saturation occurs after 160 dB when the other power constraint at $S$ is activated to prevent harmful interference at $U$. Moreover, this second saturation perfectly matches the approximated outage curve. In {\it Case~2}, the system performance (given by a brown short dash-dot line) follows the previous outage pattern until it comes to the point of 150 dB when no second saturation region appears since $S$ is not restricted by Eq. \eqref{relay_power}. Consequently, the outage probability increases and tends to 0. In the next scenario, {\it i.e.}, {\it Case~3}, the outage performance (presented by a red short dash line) of the system improves as the transmit SNR increases until the power constraint at $S$ individually determines an outage event in the network. On the other hand, {\it Case~4}, when $S$ and $R$ do not interfere with the PN, the outage performance improves as the transmit SNR increases without any saturation. 
\begin{remark}
	{\it Cases~3} and {\it 4} can be treated as the asymptotic benchmarks useful for investigation of extreme cases of the considered system model without considering the presence of the VLC communication link between EVs.
\end{remark}
In Fig. \ref{res:OP1}, we also demonstrate the outage performance for $d_{RD}=20$ m, and one can observe a performance gap related to the different value of $d_{RD}$. Additionally, we plot the outage performance of the RF-based CR network mimicking {\it Case~1}. As shown, the corresponding system performance is poor, as expected. One can notice that, in {\it Case~1}, both curves obtained for different values of $d_{RD}$ experience the first saturation at the same transmit SNR value (around 50 dB). However, the saturation regions related to the different distances between $R$ and $D$ are not the same due to the corresponding path losses. At the same time, both curves have the identical levels of the second saturation region. With this in mind, we note that the outage reduction gain provided by the proposed system model is more sufficient as the $R$-to-$D$ distance becomes bigger when the final outage saturation is determined by the ITC at $U$. 

Next, in our discussion, we will consider only {\it Case~1} since it represents the most practical and sophisticated network scenario when both secondary transmitters have the CR-based power restrictions and we can quantitatively estimate the impact of the VLC communication. 

\begin{figure}[!t]
	\centering
	\includegraphics[width=1.0\columnwidth]{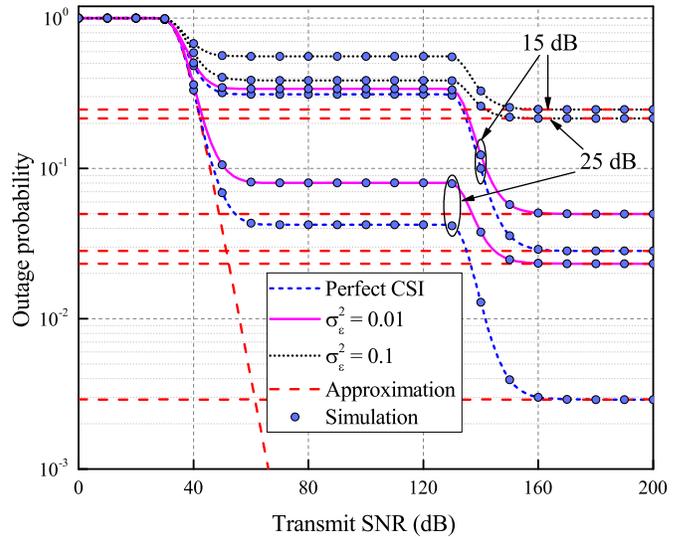}
	\caption{The outage probability versus the transmit SNR for different CSI scenarios, when $v = 2$, $I_U = \{15; 25\}$ dB, and $h_L=-65$ dB.}
	\label{res:OP11}
\end{figure}

In Fig. \ref{res:OP11}, we plot the outage performance for different CSI scenarios given by $\sigma_{\epsilon}^2 = \{0;0.01;0.1\}$, $I_{U} = \{15; 25\}$ dB and $h_L=-65$ dB. We can notice that the OP curves for the perfect CSI scenario always outperform the other CSI scenarios, as expected. Moreover, it is worthwhile pointing out that the curves related to the imperfect CSI scenarios ($\sigma_{\epsilon}^2 = 0.01$ and $\sigma_{\epsilon}^2 = 0.1$) follow the same behavior as the perfect CSI scenario, {\it i.e.}, both the saturation and further performance enhancement occur. However, they do not approach the ideal case and experience another saturation which is now caused by the CSI errors. These saturation curves coincide with the high SNR approximations of the OP performance given by Eq. \eqref{outageY_Case1}. The same behavior can be noticed for the case when the ITC at $U$ is given by $I_U = 25$ dB and the outage improves as $U$ becomes more robust to the interference level. 

\begin{figure}[!t]
	\centering
	\includegraphics[width=1.0\columnwidth]{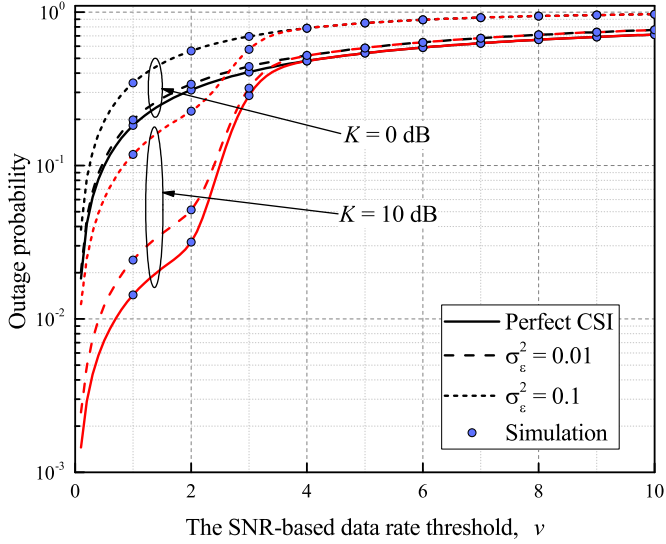}
	\caption{The outage probability versus the SNR-based data rate threshold, $v$, for $h_L = -65$ dB, $K = \{0; 10\}$ dB, and $\xi = 120$ dB.}
	\label{res:OP2}
\end{figure}
Next, in Fig. \ref{res:OP2}, we plot the outage performance versus the SNR-based data rate threshold $v$ for various CSI scenarios and different values of $K$, {\it i.e.}, 0 dB and 10 dB. As it can be noticed, the OP curves representing the perfect CSI case indicate that the system experiences a satisfactory level of network reliability, and an additional increase in $K$ sufficiently improves the OP performance. On the other hand, the network reliability severely degrades as the CSI mismatch increases, and further increase in the LOS-component power ({\it e.g.}, $K = 10$ dB) results in slight improvement of the outage metric, {\it i.e.}, this effect is less evident as the CSI quality becomes worse. Finally, we note that the identical CSI cases coincide with each other which can be explained by the fact that, after certain threshold ({\it e.g.}, in our case, it is 4 for $\xi = 120$ dB), additional power allocated to the LOS-component ({\it i.e.}, Rician fading) does not provide the outage improvement as the power constraint at $S$ gets activated which, in turn, leads to the second saturation region (see Fig. \ref{res:OP1}) due to the DF mode implemented in the considered system model.

\begin{figure}[!t]
	\centering
	\includegraphics[width=1.0\columnwidth]{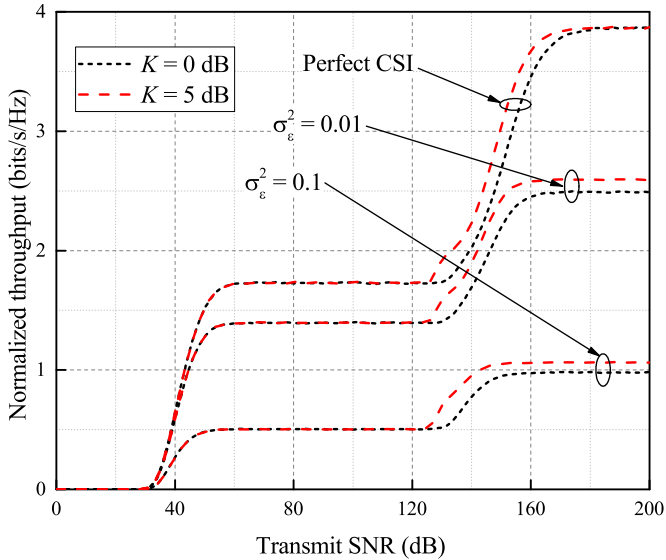}
	\caption{The normalized throughput versus the transmit SNR for $K = \{0;5\}$ dB, $v = 2$, $\xi = 120$ dB, and $I_{U} = 20$ dB.}
	\label{res:throughput}
\end{figure}
Fig. \ref{res:throughput} presents the results on the achievable throughput (normalized per Hz) which is obtained from Eq. \eqref{throuput} for different values of $K$ and CSI scenarios when $I_{U} = 20$ dB. It is important to point out that the throughput results also support the findings related to Fig. \ref{res:OP1} where one can observe the saturation regions imposed by the CR transmission which is followed by the performance improvement ({\it e.g.}, significant and small enhancements correspond to the perfect and imperfect CSI scenarios, respectively). All CSI scenarios also result in the two-stage saturation and the achievable throughput is mainly determined by the respective CSI quality. Note that, as in Fig. \ref{res:OP2}, the relative increase of the LOS-based power provides insufficient throughput improvement. Finally, another useful insight is that the quality of the CSI estimates has a stronger impact on the outage performance than the power level received from the LOS path. This can be explained by the fact that the increased power levels related to higher $K$ will be directly dispersed with respect to the CSI error variance into the desired and interference terms in the SNR expressions. For instance, more power is associated with the interference term as the CSI quality becomes worse.


\begin{figure}[!t]
	\centering
	\includegraphics[width=1.0\columnwidth]{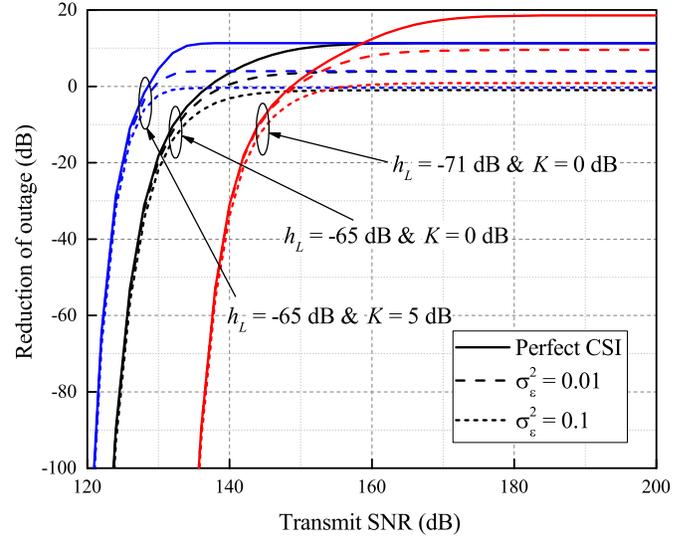}
	\caption{The reduction of outage versus the transmit SNR for different CSI scenarios when $v = 2$ and $I_{U} = 25$ dB.}
	\label{res:reduction}
\end{figure}

Next, in Fig. \ref{res:reduction}, we demonstrate some results on the reduction of outage achievable by the proposed technique for different values of $d_{RD}$ and $K$. In order to quantify this improvement, we plot the reduction of outage as $10 \log_{10}\frac{\Delta_{\rm out}}{P_{\rm out}}$ (in dB). One can note that the outage reduction for $K=5$ dB occurs earlier than the one corresponding to $K = 0$ dB. This is explained by the fact that more power is allocated to the LOS-component of the VLC channel, however, at 150 dB, both lines match each other at 150 dB since the power restriction at $S$ allows further improvement. Next, we consider the other case when $d_{RD} = 20$ m (red color), where the system starts experiencing the outage reduction gain later, around 135 dB. This happens due to the proximity between $R$ and $D$ nodes. At the same time, the outage gain is bigger compared to the previous one. It can also be observed that the reduction of outage decreases as the CSI quality degrades. Additionally, it is worthwhile noting that the imperfect CSI cases can be characterized by a fixed outage improvement because of the prevailing impact of the channel estimation errors. Another meaningful insight is that the higher the curves are depicted in Fig. \ref{res:reduction}, the more outage reduction the system achieves.

\section{Conclusion}
\label{sec:conclusion}
In this paper, we investigated the performance of the several outdoor cognitive electric vehicular network cases over mixed RF/VLC channels for different CSI scenarios. We derived novel closed-form expressions for the outage performance and analyzed their asymptotic behavior. We showed that the CSI quality effect dominates over the effect of power arrived from the LOS path. Moreover, we evaluated the achievable throughput of the system under consideration. Additionally, we quantified the outage reduction achieved by means of the mixed VLC/RF channels. Finally, we demonstrated that the derived analytical results are in agreement with the Monte Carlo simulations.

\begin{appendices}

\section{A conventional CR-based EV network over RF channels}
\label{App:1}
In this section, we derive the closed-form expressions of outage performance for different {\it Cases} in the CR-based vehicular network over RF channels. This implies that the established communication sessions do not have the advantage of availability of the VLC communication link between $R$ and $D$, {\it i.e.}, outage performance does not improve when the ITC at $U$ is reached.

\begin{align}
	\label{rf-rf_Case1}
	P^{{\it Case~}{\rm 1:~RF/RF}}_{\rm out}(v) &= 1 - 
	\exp\left(- \tilde{\lambda}_{SR} v \left(\sigma_{\epsilon}^2 + \frac{d_{SR}^{\tau} \sigma_R^2 }{P_S} \right) \right) \nonumber\\
	&~~~\times \exp\left(- \tilde{\lambda}_{RD} v \left(\sigma_{\epsilon}^2 + \frac{ d_{RD}^{\tau} \sigma_{D}^2}{P_R} \right) \right) \nonumber \\
	&~~~\times   
	\left[
	1 - \left(
	1 - \Sigma_S
	\right) \exp\left( - \lambda_{SU} \frac{K_S}{P_S} \right)  
	\right] \nonumber\\
	\times &
	\left[
	1 - \left(1 - \Sigma_R
	\right) \exp\left(- \lambda_{RU} \frac{K_R}{P_R}\right)
	\right]
\end{align}

\begin{align}
	\label{rf-rf_Case2}
	P^{{\it Case~}{\rm 2:~RF/RF}}_{\rm out}(v) &= 1 - 
	\exp\left(- \tilde{\lambda}_{SR} v \left(\sigma_{\epsilon}^2 + \frac{d_{SR}^{\tau} \sigma_R^2 }{P_S} \right) \right) \nonumber
	\\
	&~~~\times   
	\exp\left(- \tilde{\lambda}_{RD} v \left(\sigma_{\epsilon}^2 + \frac{ d_{RD}^{\tau} \sigma_{D}^2}{P_R} \right) \right) \nonumber \\
	\times & \left[
	1 - \left(1 - \Sigma_R
	\right) \exp\left(- \lambda_{RU} \frac{K_R}{P_R}\right)
	\right]
\end{align}
\begin{align}
	\label{rf-rf_Case3}
	P^{{\it Case~}{\rm 3:~RF/RF}}_{\rm out}(v) &= 1 - 
	\exp\left(- \tilde{\lambda}_{SR} v \left(\sigma_{\epsilon}^2 + \frac{d_{SR}^{\tau} \sigma_R^2 }{P_S} \right) \right) \nonumber \\
	&~~~\times \left[
	1 - \left(
	1 - \Sigma_S
	\right) \exp\left( - \lambda_{SU} \frac{K_S}{P_S} \right)  
	\right] \nonumber
	\\
	\times &   
	\exp\left(- \tilde{\lambda}_{RD} v \left(\sigma_{\epsilon}^2 + \frac{ d_{RD}^{\tau} \sigma_{D}^2}{P_R} \right) \right)
\end{align}
{\allowdisplaybreaks
\begin{align}
	\label{rf-rf_Case4}
	P^{{\it Case~}{\rm 4:~RF/RF}}_{\rm out}(v) &= 1 - 
	\exp\left(- \tilde{\lambda}_{SR} v \left(\sigma_{\epsilon}^2 + \frac{d_{SR}^{\tau} \sigma_R^2 }{P_S} \right) \right) \nonumber \\
	\times &
	\exp\left(- \tilde{\lambda}_{RD} v \left(\sigma_{\epsilon}^2 + \frac{ d_{RD}^{\tau} \sigma_{D}^2}{P_R} \right) \right)
\end{align}}

\section{Outage reduction}
\label{App:2}
In this section, we present the closed-form expressions of asymptotic outage reduction for different {\it Cases}. As we can see, in the case of perfect CSI and high SNR, {\it Cases} 1 and 2 are advantageous compared to {\it Cases} 3 and 4, since the former ones obtain non-zero outage reduction. This can be explained by the fact that mixed RF/VLC channel is established between $R$ and $D$, and the power restriction imposed at $R$ ({\it i.e.}, {\it Cases} 1 and 2) enables the system to switch the communication session from the RF band to the VLC one. 
\begin{align}
	\label{reduction_Case1_asymp}
	\Delta_{\rm out}^{Case~{\rm 1:~asymp}}(v) &= 
	\left|
	-
	\exp\left(- \tilde{\lambda}_{SR} v \sigma_{\epsilon}^2 \right) \left[1 - \left(
	1 - \Sigma_S
	\right) \right. \right. \nonumber \\
	&\hspace{-0.5cm}\left. 
	\times 
	\exp\left( - \lambda_{SU} \frac{K_S}{P_S} \right)  \right] 
	Q_1 \left(\frac{s}{\sqrt{N}}, \sqrt{\frac{v  \sigma_{\epsilon} ^2  }{N} }\right)
	 \nonumber \\
	&~~~ \times 
	\left[ 
	1-
	\exp\left(- \tilde{\lambda}_{RD} v \sigma_{\epsilon}^2  \right) 
	\right. \nonumber\\
	&\hspace{-0.8cm} \left.\left.
	\times
	\left(
	1 - \left(1 - \Sigma_R
	\right) \exp\left(- \lambda_{RU} \frac{K_R}{P_R}\right)
	\right)  
	\right]
	\right| \\
	&
	\underset{\sigma_{\epsilon}^2 \to 0}{\stackrel{(a)}{=}}
	\left|
	-
	\left(1 - \Sigma_R
	\right) \exp\left(- \lambda_{RU} \frac{K_R}{P_R}\right)
	\right. \nonumber\\
	&\hspace{-0.5cm}\left. \times
	\left[1 - \left(
	1 - \Sigma_S
	\right) \exp\left( - \lambda_{SU} \frac{K_S}{P_S} \right)  \right] 
	\right|
\end{align}
\begin{align}
	\label{reduction_Case2_asymp}
	\Delta_{\rm out}^{Case~{\rm 2:~asymp}}(v) &= 
	\left|
	-
	\exp\left(- \tilde{\lambda}_{SR} v \sigma_{\epsilon}^2 \right) 
	Q_1 \left(\frac{s}{\sqrt{N}}, \sqrt{\frac{v  \sigma_{\epsilon} ^2  }{N} }\right)
	\right. \nonumber\\
	&~~~ \times
	\left[ 
	1-
	\exp\left(- \tilde{\lambda}_{RD} v \sigma_{\epsilon}^2  \right) \right. \nonumber \\
	&\hspace{-0.8cm}
	\left.\left.\times
	\left(
	1 - \left(1 - \Sigma_R
	\right) \exp\left(- \lambda_{RU} \frac{K_R}{P_R}\right)
	\right)  
	\right]
	\right| \\
	&
	\underset{\sigma_{\epsilon}^2 \to 0}{\stackrel{(a)}{=}}
	\left|
	- \left(1 - \Sigma_R
	\right) \exp\left(- \lambda_{RU} \frac{K_R}{P_R}\right)
	\right| \nonumber\\
	&
	=
	\left|
	- \left(1 - \left(1 + \frac{\tilde{\lambda}_{RD} v d_{RD}^{\tau} \sigma_{D}^2 }{\lambda_{RU} K_R} \right)^{-1}
	\right) \right. \nonumber \\
	&~~~\left.\times
	\exp\left(- \lambda_{RU} \frac{K_R}{P_R}\right)
	\right|
\end{align}
\begin{align}
	\label{reduction_Case3_asymp}
	\Delta_{\rm out}^{Case~{\rm 3:~asymp}}(v) &= 
	\left|
	-
	\exp\left(- \tilde{\lambda}_{SR} v \sigma_{\epsilon}^2 \right) 
	\left[1 - \left(
	1 - \Sigma_S
	\right) \right. \right. \nonumber \\
	&\hspace{-0.7cm}
	\left. \times 
	\exp\left( - \lambda_{SU} \frac{K_S}{P_S} \right)  \right] 
	Q_1 \left(\frac{s}{\sqrt{N}}, \sqrt{\frac{v  \sigma_{\epsilon} ^2  }{N} }\right)
	 \nonumber \\
	&~~~\left. \times
	\left[ 
	1-
	\exp\left(- \tilde{\lambda}_{RD} v \sigma_{\epsilon}^2  \right) 
	\right]
	\right| \\
	&
	\underset{\sigma_{\epsilon}^2 \to 0}{\stackrel{(a)}{\approx}}
	0
\end{align}
{\allowdisplaybreaks
\begin{align}
	\label{reduction_Case4_asymp}
	\Delta_{\rm out}^{Case~{\rm 4:~asymp}}(v) &= 
	\left|
	-
	Q_1 \left(\frac{s}{\sqrt{N}}, \sqrt{\frac{v  \sigma_{\epsilon} ^2  }{N} }\right)
	\right.
	\nonumber\\
	&\hspace{-1.4cm} 
	\left.\times
	\exp\left(- \tilde{\lambda}_{SR} v \sigma_{\epsilon}^2 \right)  
	\left[ 
	1-
	\exp\left(- \tilde{\lambda}_{RD} v \sigma_{\epsilon}^2  \right) 
	\right]
	\right| \\
	&
	\underset{\sigma_{\epsilon}^2 \to 0}{\stackrel{(a)}{\approx}}
	0
\end{align}}
\end{appendices}

\balance

\end{document}